\documentclass{emulateapj}
\usepackage{apjfonts}
\usepackage{amssymb}
\usepackage{amsmath}
\usepackage{amsbsy}
\usepackage{epsfig}

\begin{document}

\title{The Sensitivity of First Generation Epoch of Reionization Observatories and Their Potential for Differentiating Theoretical Power Spectra}

\author{Judd D. Bowman\altaffilmark{1,2}, Miguel F. Morales\altaffilmark{1}, Jacqueline N. Hewitt\altaffilmark{1,2}}

\email{jdbowman@mit.edu; mmorales@space.mit.edu;
jhewitt@space.mit.edu}

\altaffiltext{1}{MIT Kavli Institute for Astrophysics and Space
Research, 77 Massachusetts Ave., Cambridge, MA 02139}

\altaffiltext{2}{MIT Department of Physics}


\begin{abstract}
Statistical observations of the epoch of reionization (EOR) power
spectrum provide a rich data set for understanding the transition
from the cosmic ``dark ages'' to the ionized universe we see today.
EOR observations have become an active area of experimental
cosmology, and three first generation observatories---MWA, PAST, and
LOFAR---are currently under development. In this paper we provide the
first quantitative calculation of the three dimensional power
spectrum sensitivity, incorporating the design parameters of a
planned array. This calculation is then used to explore the
constraints these first generation observations can place on the EOR
power spectrum.  The results demonstrate the potential of upcoming
power spectrum observations to constrain theories of structure
formation and reionization.

\end{abstract}


\section{Introduction}
\label{sec_introduction}

The reionization history of the universe provides an important tool
for understanding the epoch of structure formation and the appearance
of the first luminous objects. Existing experimental clues about this
period are confusing, however. The large Thompson scattering measured
by the WMAP satellite implies that reionization occurred by redshift
$z\approx15$ \citep{2003ApJS..148..161K}, whereas quasar absorption
line studies show significant neutral hydrogen at redshift
$z\approx6$ \citep{2001ApJ...560L...5D, 2001AJ....122.2850B,
2003AJ....125.1649F, 2004Natur.427..815W}. Reconciling these
measurements seems to require a fairly complicated reionization
history \citep{2003ApJ...595....1H, 2003ApJ...591...12C,
2003MNRAS.344..607S, 2004ApJ...604..484M}.


Directly observing the process of structure formation during the
epoch of reionization (EOR) would resolve these uncertainties. As the
primordial hydrogen cools and later reheats, density contrasts are
revealed as fluctuations in the brightness temperature of the
redshifted 21 cm neutral hydrogen line \citep{1972A&A....20..189S,
1979MNRAS.188..791H, 1990MNRAS.247..510S, 1994ApJ...427...25S,
2004PhRvL..92u1301L, 2005ApJ...626....1B}. Additionally, as the first
luminous objects ionize their surroundings bubbles appear in the
diffuse 21 cm emission \citep{1997ApJ...475..429M,
2000ApJ...528..597T, 2003ApJ...596....1C, 2004ApJ...608..622Z,
2004ApJ...613...16F}.

The neutral hydrogen emission from this period appears in the low
radio frequencies as faint fluctuations between $75 \rightarrow 200$
MHz (for redshifts $18 \rightarrow 6$). Directly imaging the
fluctuations will require the sensitivity of the Square Kilometer
Array \citep{2004NewAR..48.1039F}, but statistical observations of
the fluctuation power spectrum should be obtainable with the much
smaller first generation EOR observatories currently under
construction. Measuring the power spectrum and its evolution would
provide a wealth of information about structure formation and the
fundamental astrophysics behind reionization.  At high redshift,
observations would explore structure formation in the linear gravity
regime and constrain the properties of dark matter
\citep{2004PhRvL..92u1301L, 2004NewA....9..417P}, and at lower
redshifts they would probe reionization and follow the emergence and
properties of the first luminous objects \citep{2000ApJ...528..597T,
2003ApJ...593..616S, 2004ApJ...613...16F, 2004ApJ...613....1F,
2005ApJ...624..491I}.

Measuring the EOR power spectrum builds on the statistical techniques
developed for analyzing the cosmic microwave background (CMB)
anisotropy. Unlike the CMB, however, the EOR signal is fully
three-dimensional since the frequency of the redshifted 21 cm line
maps to the line-of-sight distance. Recent efforts  have shown how
the three-dimensional nature of the EOR signal can be used to
increase instrumental sensitivity
 and mitigate against foreground
contamination \citep{2004ApJ...615....7M,2004ApJ...608..622Z,
2005MNRAS.356.1519B}.

In this paper, we utilize the formalism of
\citet{2004ApJ...615....7M} and \citet{2005ApJ...619..678M} with the
parameters of a planned array to provide the first quantitative
calculation of the three dimensional power spectrum sensitivity of
forthcoming observations. The calculations include realistic
observational parameters such as antenna layout, field of view, and
antenna temperature, and provide a fiducial mark for the capabilities
of the first generation EOR observations.

We begin in Section 2 by detailing the observational parameters and
techniques used for our calculation. Section 3 then presents the
results, analyzing the EOR sensitivity as a function of length scale,
redshift, and global ionization fraction.  We conclude in Section 4
with a discussion of the additional experimental issues which must be
confronted in a realistic array.

\section{First Generation EOR Experiments}

The first generation of radio observatories targeting the EOR power
spectrum consists of three instruments that are currently under
development and should be operational by the end of the decade. The
instruments are PAST, MWA, and LOFAR.

The Primeval Structure Telescope (PAST) is a project of the National
Astronomical Observatories, Chinese Academy of Sciences. PAST is
located in the radio quiet Ulastai valley in northwestern China, and
will consist of 10,000 single-polarization yagis observing the north
celestial pole. PAST will be the first EOR array to begin acquiring
data. \citep[submitted]{Pen_PAST}

The Mileura Wide-field Array (MWA) is a collaboration of the
Massachusetts Institute of Technology, Harvard--Smithsonian Center
for Astrophysics, the Australian University Consortium, the
Australian National Telescope Facility, and the Western Australian
government. The MWA will be located at the Mileura radio quiet site
in the Western Australian desert. The low frequency portion of the
array will consist of 8,000 dual-polarization dipoles and feature
very high spatial dynamic range and calibration precision, including
polarization. \citep[in press]{Salah_MWA}

The Low Frequency Array (LOFAR) is being built by ASTRON in the
Netherlands, and will consist of 72,000 dual polarization dipoles
operating at EOR frequencies. While the radio interference
environment of the Netherlands is challenging, LOFAR will have by far
the largest collecting area of the first generation EOR arrays.
\citep{2004P&SS...52.1343K}

\subsection{The Fiducial Observation}

To calculate accurately the sensitivity of an EOR measurement, we
need to specify both the details of the instrument and the observing
strategy.  For this paper we wish to choose a set of realistic
observation parameters so that we can discuss the statistical
sensitivity of first-generation EOR experiments.

We have chosen our reference measurement to be a deep observation of
a single target field using an array configuration based on the MWA.
Of the first generation arrays, the MWA is intermediate in
theoretical sensitivity and has good control of systematic errors.

Many factors contribute to the instrumental response of an array.
Among the most important to consider are the field of view, angular
resolution, collecting area, antenna distribution, and bandwidth.

\begin{figure}
    \begin{center}
    \plotone{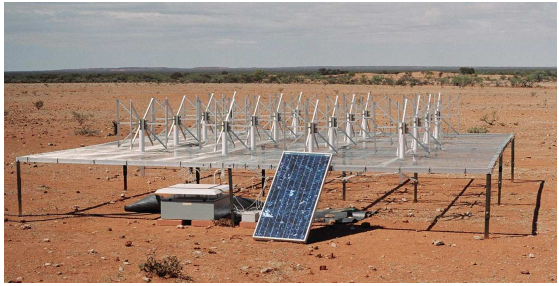}
    \caption{\label{f1}
    This photograph shows the first MWA prototype antenna being tested on
    site in Mileura.  The antenna consists of 16 crossed-dipoles in a four-by-four meter grid and is
    elevated approximately 0.5 m above the ground.  Five hundred antennas will be
    placed in a 1500 m diameter array for the low frequency band of the first-generation MWA.}
    \end{center}
    \end{figure}

For the MWA, each antenna consists of 16 crossed-dipoles in a
four-by-four meter grid as shown in Figure \ref{f1}. A beamforming
operation on the dipoles within an antenna produces a field of view
of 15-45$^\circ$ (FWHM), depending on frequency.

The reference array consists of $N=500$ antennas distributed within a
$D=1500$ m diameter circle.  The density of antennas as a function of
radius is taken to be approximately $\rho(r) \sim r^{-2}$. The
angular resolution is given by $\lambda/D$ and the collecting area by
$N~dA$, where $dA$ is the collecting area of each antenna and scales
like $ dA \sim 16 ( \lambda^2/4 )$ for wavelengths below two meters.

Although the bandwidth of the MWA is 32 MHz, we restrict the
bandwidth of our reference observation to $B=8$ MHz.  This avoids
complications introduced by cosmic evolution, yet still provides
measurements along the line-of-sight at the length scales of the
strongest fluctuations in the 21 cm signal.  It is also the minimum
bandwidth suggested by \citet{2004Natur.432..194W} to ensure
sensitivity to fluctuations during the final stage of reionization
when ionized bubbles first overlap completely.

Finally, the integration time on the target field must also be
specified. Observing for four hours per day over the course of a six
month season would result in approximately 720 hours of integration.
Conservatively rejecting half the data for non-ideal conditions
yields 360 hours of integration during the most favorable
circumstances.

The full parameter set for our fiducial observation is summarized in
Tables \ref{tab_detector} and \ref{tab_redshifts}.
    \begin{deluxetable}{lcc}
    \tablewidth{3in}
    \tablecaption
    {
        Fiducial Observation Parameters
        \label{tab_detector}
    }
    \tablecomments
    {Observation parameters used in the sensitivity analysis.
    }
    \tablehead{\colhead{Parameter} & \colhead{Value}}

    \startdata
    Array configuration, $\rho(r)$              & $\sim r^{-2}$ \\
    Array diameter, D                           & 1500 m \\
    Bandwidth, B                                & 8 MHz \\
    Frequency resolution                        & 8 kHz \\
    Integration time, t                         & 360 hours \\
    Number of antennas, N                       & 500

    \enddata

    \end{deluxetable}
    \begin{deluxetable}{lcccc}
    \tablewidth{3in}
    \tablecaption
    {
        Redshift Dependent Parameters
        \label{tab_redshifts}
    }
    \tablecomments
    { Characteristics of the fiducial observation that depend
    on frequency, and thus on redshift.  Note that the antenna
    collecting area is capped above $z=10$ due to
    self-shadowing by the antennas at low frequencies.  Also, the system temperature is dominated by sky
    noise and depends significantly on frequency.
    }
    \tablehead{\colhead{} & \colhead{$z=6$} & \colhead{$z=8$} & \colhead{$z=10$} &
    \colhead{$z=12$}}

    \startdata
    Angular resolution ($^\circ$)      & 0.06 & 0.07 & 0.09 & 0.10 \\
    Antenna collecting area, dA (m$^2$)        & 9 & 14 & 18 & 18 \\
    Antenna response scale, $\Theta$ ($^\circ$) & 19 & 31 & 38 & 43 \\
    Frequency (MHz)                 & 203 & 158 & 129 & 109 \\
    System temperature, $T_{sys}$ (K)    & 250 & 440 & 690 & 1000
    \enddata

    \end{deluxetable}

\subsection{The Data Cylinder} \label{secCylinder}

Since neutral hydrogen is optically thin to the 21 cm line, the
visibility measurements of EOR observatories inherently sample the
emission from a three dimensional volume of space at high redshift.
To good approximation, these measurements form a three dimensional
data cylinder in visibility space ($u, v, f$) due to the overall
circular shape of the array.

    \begin{deluxetable}{lcc}
    \tablewidth{3in}
    \tablecaption
    {
        Data Cylinder Dimensions for Redshift $z=8$
        \label{tab_datacylinder}
    }
    \tablecomments
    {
Dimensions of the data cylinder at redshift $z=8$ for the example
array configuration in different frames. The diameter of the cylinder
is in the plane of the first two coordinates and the depth is along
the third coordinate.
    }
    \tablehead{\colhead{Frame} & \colhead{Diameter} & \colhead{Depth (Line-of-sight)}}

    \startdata
    Cosmological Fourier ($\mathbf{k}$)     & $0.55$ Mpc$^{-1}$ & $48$ Mpc$^{-1}$ \\
    Image ($\theta_x$, $\theta_y$, $f$)     & 62$^\circ$ & 8 MHz \\
    Instrumental Fourier ($u,v,\eta$)       & $790$ $\lambda$ & $ 1.2 \cdot 10^{-4}$ Hz$^{-1}$\\
    Real                                    & $2400$ Mpc & $ 130$ Mpc \\
    Visibility ($u,v,f$)                    & $790$ $\lambda$ & $8$ MHz
    \enddata

    \end{deluxetable}

By applying Fourier transforms along one or more of the coordinates
of the data cylinder, the measurements also may be represented as
cylinders in several additional useful coordinate spaces.  These
include real space (with units of comoving Mpc), cosmological Fourier
space ($\mathbf{k} \equiv k_{1},k_{2},k_{3}$), image space
($\theta_x, \theta_y, f$), and instrumental Fourier space ($u, v,
\eta$). Each coordinate space possesses advantages for different
stages of the analysis. For example, point source foreground removal
is most conveniently accomplished in the image space, while the power
spectrum of the EOR signal has symmetries that are most easily
exploited in the Fourier space. Since space is isotropic
(rotationally invariant) the EOR signal is approximately spherically
symmetric in the Fourier space. All of the measurements from a
spherical shell are thus drawn from the same statistical ensemble and
can be averaged together to maximize the signal to noise
\citep{2004ApJ...615....7M,2005ApJ...619..678M}. The spherical
symmetry is the basis of the three dimensional statistical EOR
measurement and employed in our sensitivity calculation below.

At redshift $z=8$, the data cylinder in cosmological Fourier space
has a diameter of $0.55$ Mpc$^{-1}$ in the $k_1 k_2$-plane and spans
$48$ Mpc$^{-1}$ in the line-of-sight direction. Clearly the data
cylinder is very elongated in Fourier space, with the line-of-sight
axis responsible for the high spatial frequency contribution. Table
\ref{tab_datacylinder} lists the dimensions of the data cylinder in
additional frames.  We will see in the following sections that the
dimensions of the data cylinder play an important role in the planned
statistical measurements.

\subsection{The Instrumental Response}

We have so far outlined the details of the fiducial observation.  In
order to understand the sensitivity of the experiment, the 21 cm
power spectrum must be mapped to an instrumental response. Following
the development of \citet{2004ApJ...615....7M}, this response is
given by the convolution of the power spectrum, $P(\mathbf{k})$, with
the instrumental window function, $W(\mathbf{k})$, according to
\begin{equation}
C^I(\mathbf{k}) = \left < | \Delta I(\mathbf{k})|^2 \right > = \int
P(\mathbf{k}) | W(\mathbf{k}-\mathbf{k}') |^2 d^3k'.
\end{equation}
The window function is given by the array's spatial and frequency
response.  We approximate it with a function that is independent of
frequency within the band and that depends on $\theta$ as
\begin{equation}
    \begin{array}{cc}
        W(\theta) = \cos^2 \left ( \frac{\pi}{2} \theta / \Theta \right ), & \quad \mbox{ $\theta < \Theta$} \\
    \end{array}
\end{equation}
where $\Theta$ is the antenna response scale size and $W(\theta)$ is
zero outside the defined region. In general the Fourier space window
function, $W(\mathbf{k})$, is a very sharply peaked function. At
redshift $z=8$, the width of $W(\mathbf{k})$ for the example array
configuration is approximately 0.003 Mpc$^{-1}$ in the $k_1
k_2$-plane and 0.05 Mpc$^{-1}$ along the $k_3$-direction.

Inherent in every radio observation is thermal noise.  The
contribution of this noise can be estimated by dividing the Fourier
space into a large number of independent cells, where the size of
each cell is given by the window function, $W(\mathbf{k})$. The
thermal noise per independent cell can then be approximated using
\citep[His Equation 11]{2005ApJ...619..678M}
\begin{equation}
\label{eqn_noise} \left [C_{ij}^{N}(\mathbf{u})\right ]_{\rm rms}=
2\left (\frac{ 2 k_{B}T_{sys}}{\epsilon\, dA\,d\eta}\right
)^{2}\frac{1}{B\,\bar{n}(\mathbf{u})\,t}\delta_{ij},
\end{equation}
where $dA$ is the physical antenna area, $d\eta$ is the inverse of
the total bandwidth, $k_B$ is Boltzmann's constant, $T_{sys}$ is the
total system temperature, $\epsilon$ is the efficiency, $B$ is the
total bandwidth, $\overline{n}$ is the time average number of
baselines in an observing cell, $\mathbf{u}\equiv
\sqrt{u^{2}+v^{2}}$, and $t$ is the total observation time.

The thermal noise is approximately independent of frequency within
the observing band and has a cylindrical symmetry in Fourier space.
The measured value of the power spectrum, however, is an average over
spherical shells in Fourier space.  Thus, to calculate the
uncertainty due to thermal noise in the measurement, we must average
the uncertainty contributions from all of the independent cells in
spherical shells. The result is that the uncertainty at a given
length scale has a somewhat complicated dependence on thermal noise,
and therefore, on the antenna distribution of the array.

\begin{figure}
    \begin{center}
    \plotone{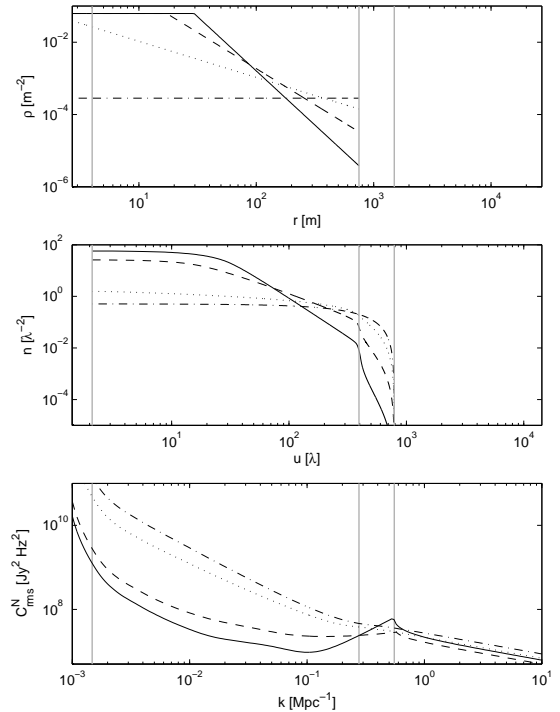}
    \end{center}
    \caption{ \label{f2} Four model antenna configurations (top), the corresponding densities
    of visibility measurements (middle), and 1-$\sigma$ uncertainties in
    the measured power spectrum due to thermal noise (bottom).  The
    antenna configurations are characterized by power-law density
    profiles, $\rho(r)$ $\sim$ $r^{-3}$ (solid), $r^{-2}$ (dash),
    $r^{-1}$ (dot), and $r^{-0}$ (dash-dot). The abscissas of the three
    panels are aligned so that the $r$, $u$, and $k$ coordinates
    correspond. The vertical gray bars represent the $4$ m width of an
    antenna (far left), the $750$ m radius of the array (middle) and the
    $1500$ m maximum baseline in the $uv$-plane (right). The
    uncertainties in the bottom panel extend beyond the bounds of the
    maximum baseline due to the elongated $k_3$ axis, as described in
    Section \ref{secCylinder} }
    \end{figure}

The top panel of Figure \ref{f2} shows four candidate antenna
profiles for the MWA. All of the distributions contain 500 antennas
and respect the minimum physically possible antenna spacing (which is
responsible for the density plateaus as $r \rightarrow 0$). The four
distributions are labelled by the antenna density profiles at large
radii: $\rho(r) \sim$ $r^0$, $r^{-1}$, $r^{-2}$, and $r^{-3}$.

The second panel shows the density of visibility measurements in the
$uv$-plane for each of the antenna distributions.  The density is
related to the average number of measurements per cell,
$\bar{n}(\mathbf{u})$, in Equation \ref{eqn_noise} by the cell size.

The third panel of Figure \ref{f2} gives the uncertainty due to
thermal noise per spherical logarithmic shell in cosmological Fourier
space with five bins per decade. Two limiting regimes are evident in
this plot. For small shells in Fourier space, the noise depends
strongly on the density profile since only visibility measurements
from baselines smaller than a shell's radius contribute. Increasing
the steepness of the antenna profile condenses the visibility
measurements toward the origin of the $k_1 k_2$-plane, reducing the
noise in small shells. In the other limiting case, shells in Fourier
space that have extended beyond the radius of the largest baseline in
the $k_1 k_2$-plane ($k \gtrsim 0.5$ Mpc$^{-1}$) include information
from every antenna, and the density profile has little effect on the
noise.

Second order effects such as the distinction between coherent
integration of visibility measurements within a cell and the
incoherent averaging of independent cells leads to the small
differences seen at large $k$. Between these limiting cases, the
interaction of the cylindrical symmetry of the thermal noise with the
spherical symmetry and logarithmic widths of the shells causes a more
complicated behavior. At length scales comparable to the diameter of
the data cylinder, where visibility measurements are sparse over much
of a shell, the uncertainty even has a local maximum for the
centrally condensed arrays.

It is clear from Figure \ref{f2} that the sensitivity of the
statistical EOR measurement is tightly linked to the array antenna
distribution.  The difference in uncertainty between the uniform
distribution and the steeper power-law distributions is as much as
two orders of magnitude, depending on the length scale of interest.
As indicated in Table \ref{tab_detector}, we will use the $\rho(r)
\sim r^{-2}$ antenna distribution as our reference in the remainder
of the paper.

\subsection{The Model Power Spectrum}

For our calculations we use a simple model of the redshifted 21 cm
power spectrum that has been used commonly in the literature, in
which the hydrogen in the intergalactic medium (IGM) is fully
neutral, follows the dark matter distribution, and has a spin
temperature much larger than the CMB temperature
\citep{1997ApJ...475..429M, 2000ApJ...528..597T,
2004ApJ...608..622Z}.  This is a reasonable model for the
fluctuations after the spin temperature has been heated by the
Wouthuysen-Field effect or other processes and before reionization by
the first luminous objects, though over what redshift range this may
be observed is uncertain.

This model power spectrum is computed using CMBFAST
\citep{1996ApJ...469..437S} and does not include velocity
distortions, but \citet{2005ApJ...624L..65B} have shown that
including peculiar velocities increases the signal amplitude by about
a factor of two. Distortions due to geometrical effects (such as a
scaling between the line-of-sight and sky-plane axes) and peculiar
velocities could allow separation of cosmological and astrophysical
effects and provide sensitive probes of the underlying cosmology, but
are not included in our simple model \citep{1979Natur.281..358A,
1987MNRAS.227....1K, 2005ApJ...624L..65B}.

\begin{figure}
    \begin{center}
    \plotone{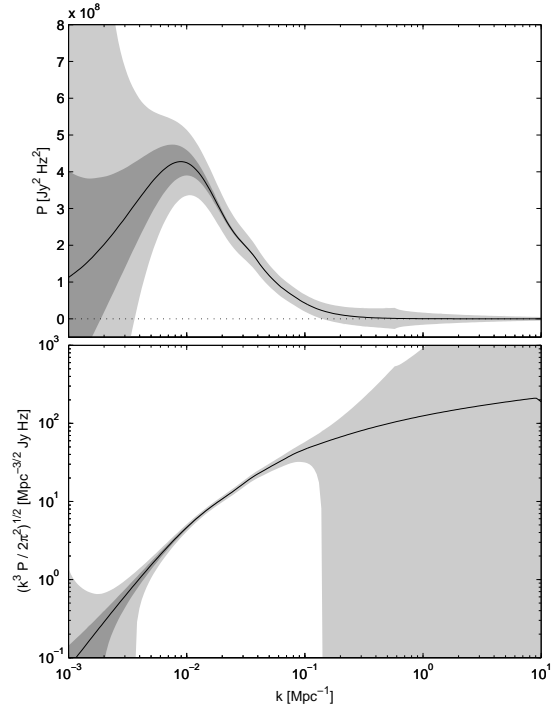}
    \end{center}
    \caption{ \label{f3}
    Combined 1-$\sigma$ uncertainty for logarithmic shells of width
    $k^{0.2}$ (equivalent to five spectral points per decade) in the
    measured power spectrum due to sample variance (dark gray) and
    combined sample and thermal variance (light gray). The instrumental
    response to the ideal neutral hydrogen power spectrum at redshift
    $z=8$ is shown in black. Logarithmic shells of width $k^{0.2}$ were
    used for bins. The details of the example observation configuration
    are specified in Tables \ref{tab_detector} and \ref{tab_redshifts}.
    In the upper panel the signal is shown in observer units (where the
    measurement errors are Gaussian), while the lower panel  uses the
    theoretical convention by changing the ordinate  to $(k^3 P / 2
    \pi^2)^{1/2}$ and plotting on log-log axes.}
    \end{figure}

Figure \ref{f3} shows the results of performing the above
calculations for the fiducial experiment using our simple neutral
hydrogen power spectrum at redshift $z=8$ with standard concordance
cosmology ($\Omega_M=0.3, \Omega_{\Lambda}=0.7, h=0.7$). The power
spectrum was convolved with the instrumental window function to
produce the solid black line, which is plotted in the observer units
Jy$^2$~Hz$^2$ (see \citet{2004ApJ...615....7M} for a discussion of
units).

Just as there is an inherent uncertainty due to thermal noise, there
is also an inherent cosmic sample variance in the observed power
spectrum. For a spherical shell in Fourier space, the sample variance
can be estimated by considering the number of independent samples in
the shell and approximating by using Gaussian statistics. This
uncertainty is plotted as the dark gray region around the ideal power
spectrum in Figure \ref{f3}, while the combined uncertainty due to
sample variance and thermal noise is added in quadrature, and the
light gray region shows the full 1-sigma uncertainty.
    \begin{figure*}
    \begin{center}
    \plotone{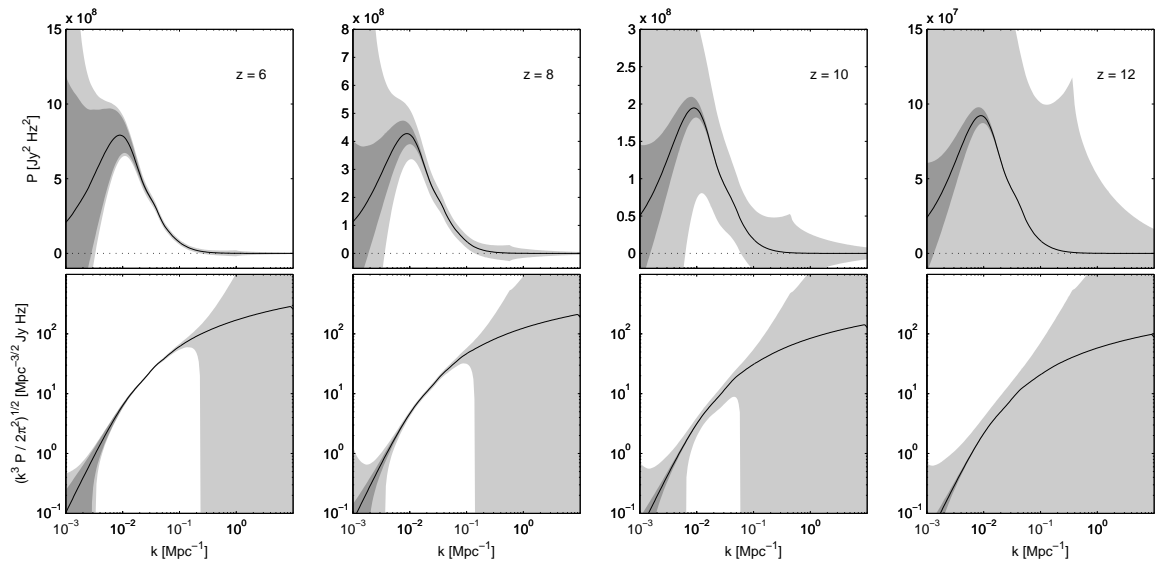}
    \end{center}
    \caption{ \label{f4}
    Same as Figure \ref{f3}, but computed for observations at four
    redshifts.  From left to right, the redshifts are $z=$ 6, 8, 10, and
    12. Note that the vertical scales are different for each of the upper
    plots. The small peaks in uncertainty at $k\approx0.5$ for $z=10$ and
    $12$ correspond to the increase in uncertainty due to thermal noise
    at length scales sampled by the largest baselines (see Figure
    \ref{f2}).}
    \end{figure*}

\section{RESULTS}
\label{sec_results}

\subsection{The Measured Power Spectrum}

Several effects of the instrumental response are contained in the
measured power spectrum shown in Figure \ref{f3}. Since the shape of
the power spectrum is smeared by convolution with the instrumental
window function, both the relative amplitude of the peak and the
distinction of the baryon bump at $k \simeq 0.04$ are slightly
reduced.  In addition, the uncertainty due to thermal noise and
cosmic sample variance increases rapidly as $k \rightarrow 0$,
creating unfavorable sensitivity on large length scales.

The field of view is a significant factor determining the uncertainty
at low $k$ since the number of measurements at these scales is
proportional to $\Theta^2$. Since the line-of-sight depth of the data
cylinder in real space is less than these length scales, the
advantages of a three-dimensional data set do not apply at low $k$.
Only spherical shells with radius $k \gtrsim 0.02$ contain
contributions from measurements with $k_3\neq0$.

The antenna distribution of the array, as we saw in Section 2, also
affects the uncertainty of the measurement. Although more condensed
arrays produced less uncertainty at large length scales (see Figure
\ref{f2}), other considerations which rely on synthesis imaging, such
as removing astrophysical foregrounds, are adversely affected by
condensing the antenna distribution and could potentially negate any
benefits from such a change.

The sensitivity decreases again for large $k$ since the power
spectrum falls off more rapidly than the thermal noise. For the
example array, the best range for constraining the measured power
spectrum at redshift $z=8$ is approximately $10^{-2} < k < 10^{-1}$
Mpc$^{-1}$.

\subsection{Redshift Range}

We can estimate the sensitivity of the array at additional redshifts
by modifying the parameters of the fiducial observation.  Four
primary characteristics of the array are frequency dependent: the
field of view (and thus the instrumental window function, $W$), the
collecting area, the angular resolution, and the system temperature,
$T_{sys}$. Table \ref{tab_redshifts} lists the values of these
parameters at four frequencies corresponding to redshifts of $z =$ 6,
8, 10, and 12.

The changes in these parameters require both the measurement
uncertainty and the instrumental response to the redshifted 21 cm
emission to be calculated for each redshift since the independent
cell size, data cylinder dimensions, and characteristic thermal noise
are directly affected. Figure \ref{f4} displays the results of such
calculations for observations at the redshifts listed in Table
\ref{tab_redshifts}. Again, the reference signal was a fully neutral
IGM and the measurement was averaged over spherical logarithmic bins
of width $k^{0.2}$.

The top panel, for redshift $z=6$, illustrates the measurement with
the greatest sensitivity.  Two factors contribute to this
performance: the amplitude of the power spectrum increases as $z
\rightarrow 0$, and the system temperature of the instrument,
primarily responsible for the uncertainty, decreases. On the other
hand, the field of view and collecting area are reduced considerably,
thus limiting the improvement in sensitivity for lower redshifts. The
net result is that the amplitude of the power spectrum increases by a
factor of $\sim 8$ between redshifts $z=12$ and $6$, while the system
temperature, dominated by galactic foreground emission, decreases by
a factor of $\sim3.5$. The sensitivity is sufficiently great at
redshift $z=6$ that the dominant source of uncertainty at large
scales is sample variance. Although it is unlikely that no ionization
will have occurred by redshift $z=6$, even significantly weaker
signals should be detectable.

There is a clear degradation of the measurement sensitivity in Figure
\ref{f4} as the redshift increases until, by redshift $z=12$, the
observation is infeasible without longer integrations or additional
collecting area.

\subsection{Sensitivity to Reionization Models}

Based on the results of the above sections, it is not unreasonable to
expect that the first generation of EOR experiments may be able to go
beyond simple detections of high redshift neutral hydrogen and
distinguish between different theoretical reionization scenarios.

\begin{figure}
    \begin{center}
    \plotone{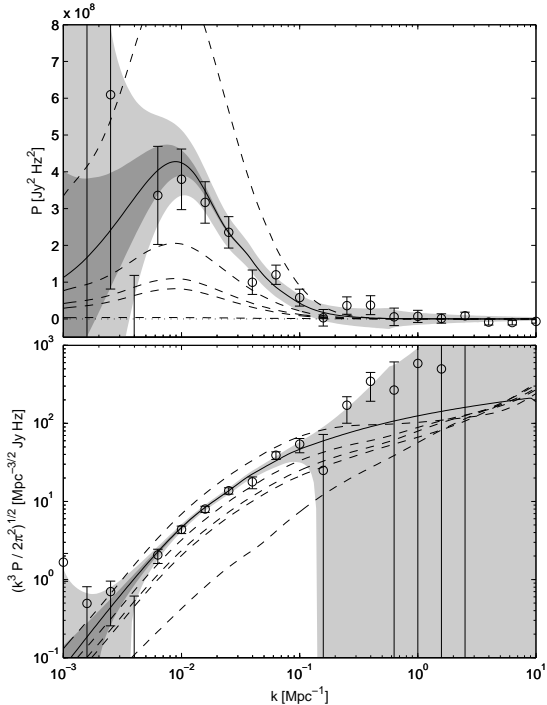}
    \end{center}
    \caption{ \label{f5}
    Same as Figure \ref{f4}, Column 2 for redshift $z=8$, but the data
    points show a simulated realization of the measured power spectrum.
    The error bars are the $1-\sigma$ uncertainties calculated from the
    thermal noise. The dashed lines are different values of the
    ionization fraction in the \citet{2004ApJ...613...16F, 2004ApJ...613....1F} models and are, from top to bottom, $x_i=0.51$, $0.0$ (solid),
    $0.43$, $0.38$, $0.25$, and $0.13$. Note that the amplitude of the
    power spectrum for the fully neutral IGM is within those of
    the other ionization fractions.  In general, the amplitudes of the
    model power spectra drop rapidly between $x_i=0.0$ and $0.13$, and
    then slowly increase with ionization fraction as large bubbles increase the contrast. }
    \end{figure}

Several theoretical reionization models have been discussed in the
literature \citep{2003ApJ...598..756S, 2004ApJ...608..622Z,
2004ApJ...613...16F, 2004ApJ...613....1F, 2005ApJ...625..575S}.
Figure \ref{f5} shows the results of a simulation based on the
example array configuration for redshift $z=8$ and includes for
comparison \citet{2004ApJ...613...16F, 2004ApJ...613....1F} models
for several power spectra with different reionization fractions. The
solid black line and shaded gray regions are the same as in Figure
\ref{f4}, Column 2. It is immediately evident from the figure that
the large changes in peak amplitude of the power spectra for these
models would be ideal for constraining the ionization fraction.

\section{EXPERIMENTAL CONSIDERATIONS}

Radio observations in the meter bands are notoriously difficult.
While this paper has focused on the theoretical sensitivity of first
generation EOR observations, it is important to remember that there
are many other observational difficulties which must be overcome by
the experiments.  In this section we briefly describe some of the
additional observational considerations.

The redshifted 21 cm radiation targeted by EOR experiments falls in
the frequency range commonly used for television, FM radio, and
satellite transmission. Both PAST and MWA have chosen very remote
locations to get away from the radio communications which are now
ubiquitous in many parts of the world, and all three experiments are
developing advanced radio frequency interference (RFI) mitigation
techniques.

Additionally, turbulence in the Earth's ionosphere refracts low
frequency radio waves. This acts much like atmospheric distortions at
optical wavelenghts, and must be corrected using techniques similar
to wide-field adaptive optics.

After RFI and ionospheric distortions have been removed from the
observations, there are still astrophysical foreground contaminants
which are five orders of magnitude brighter than the EOR emission.
Initial analysis suggested that these foregrounds---which include
synchrotron and free-free emission from the Galaxy, free-free
emission from elections in the IGM, and extended and point
sources---were an insurmountable obstacle \citep{2002ApJ...564..576D,
2003MNRAS.346..871O}, but additional studies indicate that
multi-frequency observations and statistical techniques should
provide methods to separate the foregrounds from the EOR signal
\citep{2004MNRAS.355.1053D, 2004ApJ...608..622Z, 2004ApJ...615....7M,
2005ApJ...625..575S}.

Another concern is the instrumental effects related to galactic
emission and imperfect instrumental calibration. Galactic synchrotron
radiation dominates the sky at low radio frequencies and is
responsible for the large system temperatures used in the fiducial
observation. Additionally, the radiation is Faraday rotated by the
interstellar medium and presents a bright, highly structured
polarized emission pattern across the sky.  This places very tight
constraints on the precision of the instrumental polarization
calibration.

All of these additional considerations should be manageable, but they
complicate the analysis of the EOR power spectrum in actual
observations. After removal or correction, each will produce a
characteristic residual statistical signature in the observed data
cylinder. Efforts are underway to assess the best techniques for
addressing these contaminants and to estimate their residual
statistical signatures.

\section{CONCLUSION}
\label{sec_conclusion}

Radio observations of the redshifted 21 cm emission during the epoch
of reionization are capable of providing statistical measurements of
the properties of neutral hydrogen at high redshift. These
measurements can be carried out with the relatively small
first-generation EOR instruments.  Future larger arrays, such as the
Square Kilometer Array, will be capable in principle of mapping
individual features on the sky, and larger collecting areas should
also allow high-precision measurements of the EOR power spectrum at
higher redshifts, with important implications for fundamental
cosmological measurements \citep{2004PhRvL..92u1301L}.

In this paper, we have illustrated explicitly the fundamental
uncertainties for a realistic experiment. This analysis provides a
firm foundation to motivate continued development in this field.

Measurements of the reionization history are fundamental to
understanding the evolution of the universe, and with properly
designed experiments, these measurements are feasible and may be
obtained in the very near future.

\vspace{12pt} We would like to thank Matt McQuinn, Steve Furlanetto,
and Matias Zaldarriaga for providing the theoretical reionization
power spectra used in Section \ref{sec_results}, and Brian Corey for
supplying the photograph used in Figure \ref{f1}. Support for this
work was provided by NSF grant AST-0121164 and the MIT School of
Science.



\end{document}